\documentstyle[aps,prl,preprint,floats,epsfig]{revtex} 

\def\BR{{\cal B}}

\begin{document}

\preprint{\tighten\vbox{\hbox{\hfil CLNS 01/1719}
                        \hbox{\hfil CLEO 01-03}
}}

\title{Observation of $B\to\phi K$ and $B\to\phi K^*$}

\author{CLEO Collaboration}
\date{January 18, 2001}

\maketitle
\tighten

\begin{abstract}
We have studied two-body charmless hadronic
decays of $B$ mesons into the final states
$\phi K$ and  $\phi K^*$.
Using 9.7~million $B\bar{B}$~pairs 
collected with the CLEO II detector,
we observe the decays $B^-\to\phi K^-$ and 
$B^0\to\phi K^{*0}$ with the following branching fractions: 
$\BR(B^-\to\phi K^-)=(5.5^{+2.1}_{-1.8}\pm 0.6)\times10^{-6}$ 
 and 
$\BR(B^0\to\phi K^{*0})=(11.5^{+4.5}_{-3.7}{^{+1.8}_{-1.7}}
)\times10^{-6}$. 
We also see evidence for the decays 
$B^0\to\phi K^0$ and $B^-\to\phi K^{*-}$. 
However, since the statistical significance is
not overwhelming for these modes we determine upper 
limits of 
$<12.3\times10^{-6}$ and 
$<22.5\times10^{-6}$
($90\%$ C.L.) respectively.

\end{abstract}
\newpage

{
\renewcommand{\thefootnote}{\fnsymbol{footnote}}

\begin{center}
R.~A.~Briere,$^{1}$ G.~P.~Chen,$^{1}$ T.~Ferguson,$^{1}$
H.~Vogel,$^{1}$
A.~Gritsan,$^{2}$
J.~P.~Alexander,$^{3}$ R.~Baker,$^{3}$ C.~Bebek,$^{3}$
B.~E.~Berger,$^{3}$ K.~Berkelman,$^{3}$
% ...
\end{center}
 
\small
\begin{center}
$^{1}${Carnegie Mellon University, Pittsburgh, Pennsylvania 15213}\\
$^{2}${University of Colorado, Boulder, Colorado 80309-0390}\\
$^{3}${Cornell University, Ithaca, New York 14853}\\
% ...
\end{center}

\setcounter{footnote}{0}
}
\newpage

The phenomenon of $CP$ violation can be accommodated 
in the Standard Model (SM) by a complex phase in the   
Cabibbo-Kobayashi-Maskawa (CKM) quark-mixing matrix~\cite{CKM}.
Whether this phase is the only source of $CP$ violation in nature 
remains an open experimental question. Studies of charmless 
$B$ meson decays will certainly play an important role in 
constraining the CKM matrix and testing the SM picture of 
$CP$ violation.

Flavor Changing Neutral Currents (FCNC) are forbidden 
to first order in the SM. Second order loop diagrams, 
known as penguin and box diagrams, can generate 
effective FCNC which lead to $b\to s$ transitions.
These processes are of considerable interest because 
they are sensitive to $V_{ts}$, the CKM
matrix element which will be very difficult to 
measure in direct decays of the top quark.
They are also sensitive to non-Standard Model physics~\cite{nonSM},
since charged Higgs bosons, new gauge bosons or 
supersymmetric particles can contribute via additional loop diagrams.

\begin{figure}[hbp]
\centering
\leavevmode
\epsfxsize=3.25in
\epsffile{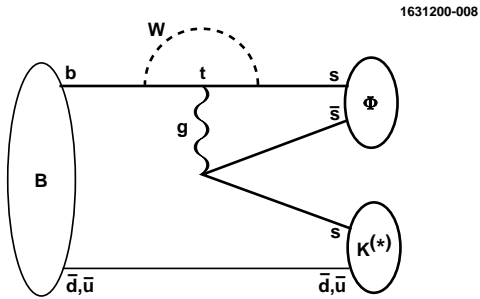}
\caption{Penguin diagram describing $B\to\phi K^{(*)}$ decays}
\label{fig:fig1}
\end{figure}

Among final states produced by the gluonic penguin, $b\to s g$,
decay modes in which a gluon splits into two strange quarks, $g\to s\bar{s}$,
play a special role since they cannot be produced by any other
$b$ decay with comparable rate, thus providing an unambiguous 
signature for the
gluonic penguin. As illustrated in Fig.~\ref{fig:fig1}, a particularly
clean final state is produced when the kaon includes the spectator
quark and no pions are emitted.
In this Letter we present the first significant 
measurements of exclusive 
charmless hadronic decays $B\to\phi K$ and $B\to\phi K^*$. 
Measurements of other charmless hadronic decay modes by CLEO 
are discussed elsewhere~\cite{bdecays}.

The data set used in this analysis was collected with the CLEO 
detector at the Cornell Electron Storage Ring (CESR).
It consists of $9.13~{\rm fb}^{-1}$ taken at the $\Upsilon$(4S)
(on-resonance), corresponding to $9.66M$ $B\bar{B}$ pairs, 
and $4.35~{\rm fb}^{-1}$ taken below $B\bar{B}$ threshold, 
used for continuum background studies.  

CLEO II is a general purpose solenoidal magnet detector~\cite{detector}. 
Cylindrical drift
chambers in a 1.5T solenoidal magnetic field measure momentum and
specific ionization ($dE/dx$) of charged particles. Photons are detected
using a 7800-crystal CsI(Tl) electromagnetic calorimeter.  In the CLEO II.V
detector configuration, the innermost chamber was replaced by a 
three-layer, double-sided silicon
vertex detector, and the gas in the main drift chamber was changed
from an argon-ethane to a helium-propane mixture. As a result of these 
modifications, the CLEO II.V portion of the data (2/3 of the total) has
improved particle identification and momentum resolution.

In the analysis presented here we search for $B$ meson decays 
by selecting $\phi$ and $K$ ($K^*$) decay candidates using 
specific criteria.
Track quality requirements are imposed on charged tracks,
and pions and kaons are identified by $dE/dx$. 
Electrons are rejected based on $dE/dx$ and the ratio of
the track momentum to the associated shower energy in the CsI
calorimeter; muons are rejected based on their penetration depth in
the instrumented steel flux return.  
The $K^0_S$ candidates are selected from 
$\pi^+\pi^-$ pairs forming well-measured secondary vertices
with invariant mass within three standard deviations
($\sigma$) of the nominal $K^0_S$ mass 
and a decay path significance of at least 3$\sigma$.
The neutral pion candidates are formed from pairs of isolated 
photon-like energy clusters in the CsI calorimeter 
with invariant mass within $-3.5$ and $+3.0$ standard deviations 
of the $\pi^0$ mass. 
The $\phi$ meson candidates have $K^+K^-$ mass
within $\pm20$ MeV/$c^2$ ($\pm 4.5 \Gamma$, $\Gamma=$ natural width) 
of the known $\phi$ mass, and 
the specific ionization of the tracks are 
consistent with the $K^+K^-$ hypothesis.
The $K^*$ candidates are reconstructed
in four modes: $K^{*0}\to K^-\pi^+$, 
$K^{*0}\to K^0\pi^0$, $K^{*-}\to K^-\pi^0$ and 
$K^{*-}\to K^0\pi^-$, and their masses lie within 
$\pm 75$ MeV/$c^2$  ($\pm 1.5\Gamma$)
of the respective known masses.
Charmless two-body $B$ decays produce the fastest secondary 
particles among all $B$ decays.
Therefore, to reduce combinatoric backgrounds 
only the fastest $\phi$ and 
the fastest $K$ ($K^*$) are used in those events with multiple 
combinations.
 
The $B$ decay candidate is identified via its invariant mass 
and its total energy.
We calculate a beam-constrained $B$ mass 
$M_B = \sqrt{E_{\rm beam}^2 - p_B^2}$, where $p_B$ is the $B$
candidate momentum and $E_{\rm beam}$ is the beam energy.
The resolution in $M_B$ is dominated by the beam energy spread. 
We define $\Delta E = E_1 + E_2 - E_{\rm beam}$, where $E_1$ and $E_2$
are the energies of the daughters of the $B$ meson candidate.
We accept events with $M_B$\ above $5.2$~GeV/$c^2$ and 
$|\Delta E|<200$~MeV.

We have studied backgrounds from $b\to c$\ decays and other $b\to u$\
and $b\to s$\ decays and find that all are negligible for the analyses
presented here. The main background arises from $e^+e^-\to q\bar q$\
(where $q=u,d,s,c$).  Such events typically exhibit a two-jet
structure and can produce high momentum back-to-back tracks in the
fiducial region, while $B\bar B$ events tend to have more spherical 
structure, since the $B$ mesons are produced nearly at rest.  
To reduce contamination from these background events, we
require the event to have $H_2/H_0 <0.6$, where $H_i$ are 
Fox-Wolfram moments~\cite{Fox}.

We extract the signal yields from unbinned,
extended maximum-likelihood fits of the preselected
on-resonance data separately for each topology 
($\phi K^-$, $\phi K^0$, 
$\phi K^{*0}_{\to K^-\pi^+}$, $\phi K^{*0}_{\to K^0\pi^0}$, 
$\phi K^{*-}_{\to K^-\pi^0}$, $\phi K^{*-}_{\to K^0\pi^-}$).
For all modes, we distinguish signal from background using
$M_B$, $\Delta E$, $|\cos\theta_{tt}|$ (the angle 
between the thrust axes of the $B$ candidate and that of 
the rest of the event), 
$|\cos\theta_B|$ (the angle between the $B$ candidate momentum 
and beam axis) and $m_\phi$ (the mass of the $\phi$ candidate). 
In addition, we include 
$|\cos\theta_h|$ (the $\phi$ helicity angle, defined as the kaon
direction in the $\phi$ rest frame) 
for $\phi K^-$ and $\phi K^0$ modes, 
$m_{K^*}$ (the mass of the $K^*$ candidate) 
for $\phi K^*$ modes, and $S_K$ (the number of standard deviations from the 
predicted $dE/dx$ value)
of $K^-$ when applicable.

In each of these fits, the likelihood of the event is 
the sum of probabilities for the 
signal and background hypotheses, with relative weights 
determined by maximizing the likelihood function $\cal L$.  
The probability of a particular hypothesis is calculated 
as a product of the probability density functions (PDFs) 
for each of the input variables.
The signal PDFs are represented by a double Gaussian 
for $M_B$ and $\Delta E$, by a Breit-Wigner function on top 
of a linear polynomial for $m_\phi$ and $m_{K^*}$, 
by $1-|\cos\theta_B|^2$ for $|\cos\theta_B|$, by a 
third order polynomial for $|\cos\theta_{tt}|$, 
and by $\cos^2\theta_h$ for $|\cos\theta_h|$. 
The background distributions for the intermediate resonance
masses are parameterized by the sum of a Breit-Wigner and a low-order
polynomial.
For $M_B$, we use an empirical shape 
($f(M_B)\propto M_B\sqrt{1-x^2}\exp[-\gamma(1-x^2)]$; 
$x=M_B/E_{\rm beam}$)~\cite{argusbackground}.
The background $\Delta E$ and $|\cos\theta_B|$ PDFs are both linear functions,
and $|\cos\theta_{tt}|$ is
parameterized by the sum of two terms 
$|\cos\theta_{tt}|^{\alpha}$ with different exponents.
The signal and background dependences of $S_K$ are 
bifurcated Gaussian functions.

The parameters for the PDFs 
are determined from off-resonance data (background) and 
from high-statistics Monte-Carlo (MC) samples (signal). 
In the signal MC data, we model $B\to\phi K^{(*)}$ as a
two-body decay, where for $\phi K^*$ we assume equal amplitudes for
longitudinal and transverse polarizations.  Dependence on
the unknown decay polarization is included in the systematic uncertainty.
We use a Geant~\cite{geant} based simulation to model the detector
response in detail.  Further details about the likelihood fit method
can be found in Ref.~\cite{bigrare}.

\begin{figure}[htbp]
\centering
\epsfxsize=3.25in
\epsffile{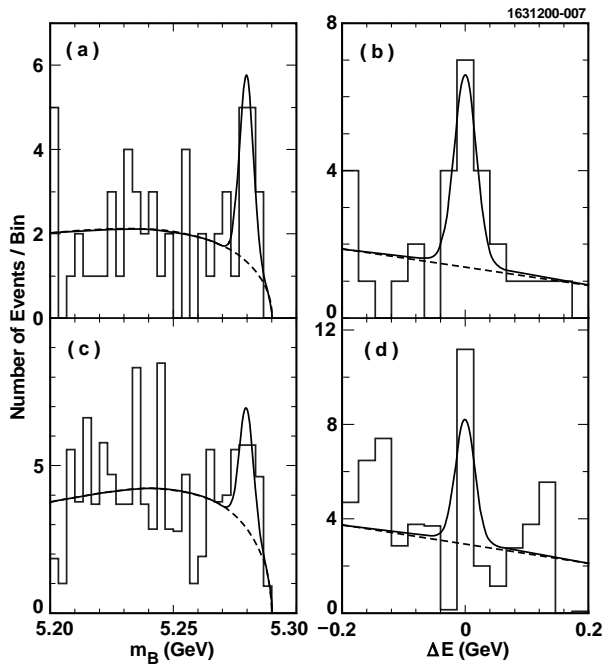}
\caption{$M$ and $\Delta E$ plots for
(a,b) $B^-\to \phi K^-$, and (c,d) $B^0\to \phi K^{*0}$
after the requirement on $R$ as described in the text. 
The projection of the total likelihood fit (solid curve)
and the continuum background component (dashed curve) are overlaid.
For $B^0\to \phi K^{*0}$, the two decay modes of $K^{*0}$ 
were weighted according to the statistical errors of the fits.
}
\label{fig:projections}
\end{figure}

To illustrate the fits, we show in Fig.~\ref{fig:projections} 
$M_B$ and $\Delta E$ projections for the modes with 
significant signals. 
Events entering these plots must satisfy a requirement 
on the signal-to-background likelihood ratios 
$R\equiv \log(P_{s\,i}/P_{b\,i})$, where $P_{s\,i}(P_{b\,i})$ 
are signal (background) likelihoods for each event $i$,
computed without $M_B$ and $\Delta E$, respectively.
This additional cut accepts about 2/3 of the preselected 
signal events in the MC sample. 

We summarize the results for all $B$ decay modes, corresponding 
submodes, and the combined modes in Table~\ref{tab:finalresults}, 
where we assume equal branching fractions for 
charged and neutral $B$ meson decays~\cite{equalbr}. 
\begin{table}[bthp]
\begin{center}
\caption{
Intermediate fitted branching fractions (${\cal B}_{fit}$), 
final branching fraction results (${\cal B}$), and theoretical 
estimates [10] are given in units of $10^{-6}$.
When the result is not statistically significant, the final result 
is quoted as a $90\%$ C.L. upper limit. The errors on branching 
fractions are statistical and systematic respectively.
Reconstruction efficiency ${\cal E}$ does not 
include branching fractions, and it is quoted for modes 
with $K^0$ assuming $K^0\to K^0_S\to\pi^+\pi^-$ decay. 
}
\begin {tabular}{l c c c c c c}
Mode& Yield & ${\cal E}(\%)$ & ${\cal B}_{fit}$ 
            & Stat. Signif. & ${\cal B}$ & Theory $\cal B$ \\
\hline
$\phi K^- $ & $14.2^{+5.5}_{-4.5}$ & 54 & $5.5^{+2.1}_{-1.8}\pm 0.6$ 
            & 5.4$\sigma$ & see ${\cal B}_{fit}$ & 0.7-16 \\
$\phi K^0 $ &  $4.2^{+2.9}_{-2.1}$ & 48 & $5.4^{+3.7}_{-2.7}\pm 0.7$ 
            & 2.9$\sigma$ & $<12.3$ & 0.7-13 \\
\hline
$\phi K$ comb. &  --- & --- & --- 
            & 6.1$\sigma$ & $5.5^{+1.8}_{-1.5}\pm 0.7$ & \\
\hline
$\phi K^{*0}$ ($K^-\pi^+$) & $12.1^{+5.3}_{-4.3}$ & 38 
            & $9.9^{+4.3}_{-3.5}{^{+1.6}_{-1.6}}$ & 4.5$\sigma$ & --- & \\
$\phi K^{*0}$ ($K^0\pi^0$) &  $5.1^{+3.9}_{-2.8}$ & 20 
            & $46.3^{+35.7}_{-26.0}{^{+5.9}_{-6.6}}$ & 2.7$\sigma$ & --- & \\
$\phi K^{*0}$ & --- & --- & $11.5^{+4.5}_{-3.7}{^{+1.8}_{-1.7}}$ 
            & 5.1$\sigma$ & see ${\cal B}_{fit}$ & 0.2-31 \\
\hline
$\phi K^{*-}$ ($K^-\pi^0$) & $3.8^{+4.1}_{-2.8}$ & 25 
            & $9.3^{+10.1}_{-7.0}{^{+1.7}_{-1.5}}$ & 1.5$\sigma$ & --- & \\
$\phi K^{*-}$ ($K^0\pi^-$) & $4.0^{+3.1}_{-2.2}$ & 32 
            & $11.4^{+9.0}_{-6.3}{^{+1.8}_{-1.8}}$ & 2.7$\sigma$ & --- & \\
$\phi K^{*-}$ & --- & --- & $10.6^{+6.4}_{-4.9}{^{+1.8}_{-1.6}}$ 
            & 3.1$\sigma$ & $<22.5$ & 0.2-31 \\
\hline
$\phi K^*$ comb. & --- & --- & --- 
                 & 5.9$\sigma$ & $11.2^{+3.6}_{-3.1}{^{+1.8}_{-1.7}}$ & \\
\end {tabular}
\label{tab:finalresults}
\end{center}
\end {table}
We combine the samples from 
multiple secondary decay channels by adding the $-2\log \cal L$ 
functions of the branching fraction. The statistical significance 
of a given signal yield is determined from the change 
in $-2\log \cal L$ when refit with the signal yield fixed 
to zero. The largest contributions to the systematic 
uncertainties come, with about equal weight, 
from uncertainties in the parametrization of the PDFs, 
decay polarization dependence,\footnote{
The $B\to\phi K^*$ decay may be longitudinally or transversely polarized. 
Assuming 100\% transverse polarization we obtain 
$\BR(B\to\phi K^*)=(13.6^{+5.3}_{-4.4})\times10^{-6}$ 
(statistical errors only) and 
$\BR(B^-\to\phi K^{*-})=(12.8^{+7.6}_{-5.9})\times10^{-6}$. 
Assuming 100\% longitudinal polarization we measure 
$\BR(B\to\phi K^*)=(9.9^{+4.2}_{-3.4})\times10^{-6}$ 
and 
$\BR(B^-\to\phi K^{*-})=(9.9^{+6.0}_{-4.6})\times10^{-6}$. 
To estimate the uncertainty due to the unknown polarization we assumed 
that any value between 100\% longitudinal and 100\% 
transverse polarization is equally likely.
} and possible background from 
other $B$ decays.

We observe a significant signal (above $5\sigma$) 
for the decays $B^-\to\phi K^-$ and $B^0\to\phi K^{*0}$. 
Since the statistical significances for the 
$B^0\to\phi K^0$ and $B^-\to\phi K^{*-}$ modes are not 
large ($2.9\sigma$ and $3.1\sigma$ respectively),  
we calculate $90\%$ confidence level 
(C.L.) upper limits (UL) by
integrating the likelihood curve to $90\%$ of its total area,
and increasing it by one unit of the systematic error.

In summary, we have measured 
$\BR(B^-\to\phi K^-)=(5.5^{+2.1}_{-1.8}\pm 0.6)\times10^{-6}$ 
 and 
$\BR(B^0\to\phi K^{*0})=(11.5^{+4.5}_{-3.7}{^{+1.8}_{-1.7}}
)\times10^{-6}$ 
each with statistical significance above $5\sigma$. 
The statistical significance of the 
$B^0\to\phi K^0$ and $B^-\to\phi K^{*-}$ signals are 
$2.9\sigma$ and $3.1\sigma$ respectively. 
The measured rates are
$\BR(B^0\to\phi K^0)= (5.4\,{^{+3.7}_{-2.7}}\pm 0.7)\times10^{-6}$ 
and 
$\BR(B^-\to\phi K^{*-})=(10.6^{+6.4}_{-4.9}{^{+1.8}_{-1.6}}
)\times10^{-6}$.
Since the statistical significance in these modes is
limited we set upper limits of 
$<12.3\times10^{-6}$ and 
$<22.5\times10^{-6}$ (at $90\%$ C.L.)
respectively.
Averaging over $B^0$ and $B^-$ 
we obtain 
$\BR(B\to\phi K)=(5.5^{+1.8}_{-1.5}\pm 0.7)\times10^{-6}$ 
($6.1\sigma$) and 
$\BR(B\to\phi K^*)=(11.2^{+3.6}_{-3.1}{^{+1.8}_{-1.7}}
)\times10^{-6}$ ($5.9\sigma$). 
The measured branching fractions lie in the range of theoretical 
predictions (see Table~\ref{tab:finalresults}). Since there is a 
considerable spread in theoretical predictions among various 
calculations, our results will help constrain model parameters. 

We gratefully acknowledge the effort of the CESR staff in providing us with
excellent luminosity and running conditions.
This work was supported by 
the National Science Foundation,
the U.S. Department of Energy,
the Research Corporation,
the Natural Sciences and Engineering Research Council of Canada, 
the Swiss National Science Foundation, 
the Texas Advanced Research Program,
and the Alexander von Humboldt Stiftung.


\begin{thebibliography}{99}

\bibitem{CKM} 
M.~Kobayashi and K.~Maskawa, Prog.\ Theor.\ Phys.\ {\bf 49},
652 (1973).

\bibitem{nonSM}
See for example:
J.L.~Hewett and J.D.~Wells, Phys.\ Rev.\ D {\bf 55}, 5549 (1997);
G.~Burdman, Phys.\ Rev.\ D {\bf 52}, 6400 (1995);
N.G.~Deshpande, K.~Panose, and J.~Trampeti\'c, 
      Phys.\ Lett.\ B {\bf 308}, 322 (1993);
W.S.~Hou, R.S.~Willey, and A.~Soni, Phys.\ Rev.\ Lett.\ {\bf 58}, 1608 (1987).

\bibitem{bdecays}
R.~Godang {\it et al.} (CLEO Collaboration),
      Phys.\ Rev.\ Lett.\ {\bf 80}, 3456 (1998);
B.~H.~Behrens {\it et al.} (CLEO Collaboration),
      Phys.\ Rev.\ Lett.\ {\bf 80}, 3710 (1998);
T.~Bergfeld {\it et al.} (CLEO Collaboration),
      Phys.\ Rev.\ Lett.\ {\bf 81}, 272 (1998);
D.~Cronin-Hennessy {\it et al.} (CLEO Collaboration),
      Phys.\ Rev.\ Lett.\ {\bf 85}, 515 (2000);
S.~J.~Richichi {\it et al.} (CLEO Collaboration),
      Phys.\ Rev.\ Lett.\ {\bf 85}, 520 (2000);
C.~P.~Jessop {\it et al.} (CLEO Collaboration),
      Phys.\ Rev.\ Lett.\ {\bf 85}, 2881 (2000).

\bibitem{detector}
Y.~Kubota {\it et al.} (CLEO Collaboration),
      Nucl.\ Instrum.\ Methods Phys.\ Res., Sec.\ A{\bf 320}, 66 (1992);
T.~S.~Hill, Nucl.\ Instrum.\ Methods Phys.\ Res., Sec.\ A{\bf 418}, 32 (1998).

\bibitem{Fox}
G.~Fox and S.~Wolfram, Phys.\ Rev.\ Lett.\ {\bf 41}, 1581 (1978).

\bibitem{argusbackground} H.~Albrecht {\it et al.} (ARGUS Collaboration),
      Phys.\ Lett.\ B {\bf 241}, 278 (1990);
H.~Albrecht {\it et al.} (ARGUS Collaboration),
      Phys.\ Lett.\ B {\bf 254}, 288 (1991).

\bibitem{geant} R.~Brun {\it et al.} CERN DD/EE/84-1.

\bibitem{bigrare}
D.~M.~Asner {\it et al.} (CLEO Collaboration),
      Phys.\ Rev.\ D {\bf 53}, 1039 (1996).

\bibitem{equalbr}
J.~P.~Alexander {\it et al.} (CLEO Collaboration),
      Report No. CLNS 00/1670, CLEO 00-7, 
      Phys.\ Rev.\ Lett.\ (to be published).

\bibitem{all}
N.~G.~ Deshpande and J.~ Trampetic, Phys.\ Rev.\ D {\bf 41}, 895 (1990);
L.-L.~Chau {\it et al.}, Phys.\ Rev.\ D {\bf 43}, 2176 (1991);
A.~ Deandrea {\it et al.}, Phys.\ Lett.\ B {\bf 318}, 549 (1993);
A.~ Deandrea {\it et al.}, Phys.\ Lett.\ B {\bf 320}, 170 (1994);
A.J.~Davies, T.~Hayashi, M.~Matsuda, and M.~Tanimoto, 
                            Phys.\ Rev.\ D {\bf 49}, 5882 (1994);
G.~Kramer, W.~F.~Palmer, and H.~Simma, 
                            Nucl.\ Phys.\ B {\bf 428}, 429 (1994);
G.~Kramer, W.~F.~Palmer, and H.~Simma, 
                            Zeit.\ Phys.\ C {\bf 66}, 429 (1995);
D.~Du and L.~Guo, Zeit.\ Phys.\ C {\bf 75}, 9 (1997).

\end{thebibliography}
\end{document}